\documentclass{PoS}

\title{Preliminary Results on the Experimental Investigation of the Structure Functions of Bound Nucleons}
\ShortTitle{Experimental Investigation of the Structure Functions of Bound Nucleons}

\author{\speaker{A. Bodek}
\\    for Jefferson Lab  experiments E02-109,  E04-001, and  E06-009.
\\Department of Physics and Astronomy, University of Rochester, Rochester, NY 14627, USA\\
        E-mail: \email{bodek@pas.rochester.edu}}

\abstract{
We present  preliminary results on an  experimental study of the  nuclear modification of the longitudinal ($\sigma_L$) and  transverse ($\sigma_T$) structure functions of nucleons bound in   nuclear targets.   The  origin of these  modifications (commonly referred as as the EMC effect)  is  not fully understood. 
Our measurements of    R= $\sigma_L / \sigma_T$ for  nuclei ($R_A$) and for deuterium ($R_D$) indicate that  nuclear modifications of the  structure functions of bound nucleons are different for the longitudinal and transverse structure functions,  and that contrary to expectation from several theoretical models,  
$R_A< R_D$.
          }

\FullConference{The XXIII International Workshop on Deep Inelastic Scattering and Related Subjects\\
		April 27 - May 1, 2015 
                Southern Methodist University
		Dallas, Texas 75275\\
		Also presented at:\\
		 The 12th Conference on the Intersection of Nuclear and Particle Physics, 
CIPANP 2015\\  May 19-24, 2015 (Vail, CO, USA) }
		
\begin{document}

In 1983,  muon (EMC\cite{emc}) and electron (MIT-SLAC\cite{mit}) scattering experiments on iron discovered that the quark distributions in the nucleon are modified when the nucleon is bound in a nucleus.  Subsequent experiments\cite{Dasu},\cite{E139} have investigated the effect for a a range of nuclei.
 The origin of these modifications  is still not fully understood.\cite{miller,JHA}.  
 
 We investigate\cite{Vahe}  these modifications  though a high precision experimental  study of both the longitudinal ($\sigma_L$) and  transverse ($\sigma_T$ ) structure functions in electron scattering on deuterium (D), carbon (C), aluminum (Al), iron (Fe),  and copper (Cu)  nuclei.  These data were taken by 
 by several different experiments (See references \cite{jupiter},\cite{E02-109},\cite{E04-001},\cite{E06-009}) which  
  cover a range in the square of the invariant momentum tansfer ($Q^2$) and energy  transfer ($\nu$)  to the nucleon, spanning hadronic  final state invariant mass $W$  ($W^2=M^2+2M\nu-Q^2$) from the resonance region (W$<$2 GeV)  to the inelastic continuum (W$>$2 GeV).

 The differential cross section for scattering of an
unpolarized charged lepton with  $E_0$, final energy
$E^{\prime}$ and scattering angle $\theta$ can be written in terms of
the structure functions ${\cal F}_1$ and ${\cal F}_2$ as:
\begin{tabbing}
$\frac{d^2\sigma}{d\Omega dE^\prime}(E_0,E^{\prime},\theta)  =
   \frac{4\alpha^2E^{\prime 2}}{Q^4} \cos^2(\theta/2)$ 
  $\times   \left[{\cal F}_2(x,Q^2)/\nu +  2 \tan^2(\theta/2) {\cal F}_1(x,Q^2)/M\right]$
\end{tabbing}
where $\alpha$ is the fine structure constant, $M$ is the nucleon
mass, $\nu=E_0-E^{\prime}$, and 
 $Q^2=4E_0E^{\prime} \sin ^2 (\theta/2)$.
Within the quark parton model,  the target mass scaling variable\cite{TM}   
$\xi_{TM}  = \frac{Q^2}
        {M\nu [1+\sqrt{1+Q^2/\nu^2}]}$
 is  the fractional momentum (parallel to the direction of the momentum transfer) carried by the struck parton in the nucleon. At large $Q^2$,  $\xi_{TM}=x$ where $x=Q^2/2M\nu$ is the
Bjorken scaling variable.

In Quantum Chromodynamics (QCD)
${\cal F}_2(x,Q^2)$  is expressed in
terms of charge weighted sums of the  fractional momentum
distributions of quarks and antiquarks in the nucleon.  The ratio
of  ${\cal F}_2$ for a nuclear target (${\cal F}_{2A}$)  to  ${\cal F}_2$ for deuterium  (${\cal F}_{2D}$)
is a function of $x$.
The  $x<0.1$ region, where ${\cal F}_{2A}/{\cal F}_{2D}<1$,  is known as the shadowing region.
The $0.1<x<0.3$ region, where  ${\cal F}_{2A}/{\cal F}_{2D}>1$, is known as the anti-shadowing
region. The   $0.3<x<0.9$ region, where ${\cal F}_{2A}/{\cal F}_{2D}<1$,  is referred to as the
region of the  "EMC Effect", and the $x>0.9$ region,  where  ${\cal F}_{2A}/{\cal F}_{2D}>1$, is dominated by Fermi motion. 
In our studies, we focus on the region of the  "EMC Effect", $0.3<x<0.9$, where the data indicate that there is softening
of the fractional momentum distribution of quarks for nucleons bound in the nucleus 

	Alternatively, one can view this scattering process 
in terms of the cross section for the
absorption of transverse $(\sigma_T)$ and longitudinal $(\sigma_L)$
virtual photons, where 
\begin{eqnarray}
\frac{d^2\sigma}{d\Omega dE^\prime} =
   \Gamma \left[\sigma_T(x,Q^2) + \epsilon \sigma_L(x,Q^2) \right].\label{req}
\end{eqnarray}   
   Here $K = \frac{2M \nu - Q^2 }{2M}$, 
   and
   $ \Gamma = \frac{\alpha K E^\prime}{ 4 \pi^2 Q^2 E_0}  \left( \frac{2}{1-\epsilon } \right)$ is
   the flux 
   and  $\epsilon = \left[ 1+2(1+\frac{Q^2}{4 M^2 x^2} ) tan^2 \frac{\theta}{2} \right] ^{-1}$
  is the degree of longitudinal polarization of the virtual photons.
 The structure functions  ${\cal F}_1$ and ${\cal F}_2$ are
proportional to $\sigma_T$,  and $[\sigma_T + \sigma_L]$, respectively. Here,    
$ {\cal R}=\sigma_L/\sigma_T$ is given by:
%
\begin{equation}
 {\cal R}(x,Q^2)
   = \frac {\sigma_L }{ \sigma_T}
   = \frac{{\cal F}_2 }{ 2x{\cal F}_1}(1+\frac{4M^2x^2 }{Q^2})-1
   = \frac{{\cal F}_L }{ 2x{\cal F}_1}
\end{equation}
  Contributions to $R$ originate
from the  perpendicular component of  the momentum of the quarks\cite{RF}  with respect
to momentum transfer vector.  A perpendicular momentum component can originated from QCD gluon emission\cite{rqcd} ($R_{QCD}$), from quark binding in the nucleon (target mass corrections\cite{TM} ($R_{TM}$),  and from  non-perturbative processes such as interactions with more than one quark (higher twist).  
Numerically, $R_{QCD}$ dominates at small $x$,  $R_{TM}$ dominates at large $x$, and higher twist effects  are important at very small $Q^2$ (since $R$ must be zero at $Q^2=0$).  
Nuclear binding can modify both the parallel and perpendicular momentum distributions of quarks,  antiquarks and gluons  of  bound nucleons thus  modifying both $F_2 $ and $R$.

Here, we report  on some preliminary results\cite{Vahe} on the  data\cite{highQ}  taken in 2007. These data were taken  
  with D, C, Al, Fe and Cu targets  for  $Q^2$ 2, 3 and 3.7 GeV$^2$.  Since quark-hadron duality has been shown to be valid\cite{duality} in this region, we can investigate  the EMC effect at large $\xi_{TM}$ by extending the studies to the resonance region.
   Incident electrons at seven different energies
provided by the Jefferson Lab accelerator  are scattered from 
a 4-cm-long liquid deuterium  target, and solid nuclear targets.
 Electrons
are detected in the Hall C
High Momentum Spectrometer (HMS) at angle settings  ranging from 
12$^0$ to 75$^0$. 
The  charge symmetric (CS) backgrounds from  the $e^+e^-$ conversion of photons
from  $\pi^0$production and its subsequent
positrons are measured in runs with reverse polarity.  Background
from eletro-produced charged pions are identified and removed
by using both a gas Cherenkov counter and an
electromagnetic calorimeter.  Events scattering
from the walls (Al)  of the cryogenic target cell are subtracted by measuring
the scattering from an  empty target replica. 
For additional details regarding
the analysis and Hall C apparatus employed
in this experiment, see Refs. \cite{Vahe}  and \cite{Christy}.

The differential cross sections are determined from the electron rates after
correcting for inefficiencies, background and radiative corrections (which include
bremsstrahlung, vertex corrections and loop diagrams standard to electron
scattering experiments).  
The cross sections are interpolated to fixed $Q^2$ values of 2, 3 and 4 GeV$^3$,
in bins of $W$. 
The  longitudinal and transverse cross sections
and structure functions are extracted from 
linear fits (equation \ref{req}) to cross section measurements
spanning a range of $\epsilon$ ($0.2<\Delta \epsilon<0.6$) for fixed values of $W$ and $Q^2$. 

  

Overall normalization errors cancel in the extraction of $R_D$. 
The overall 1.4\%  systematic error\cite{Vahe}  in the  $\epsilon$ dependence of the cross
 that originates from  $E_0$  (0.25\%),  beam charge (0.3\%), efficiencies (0.35\%), 
  CS background (0.1-0.4\%), acceptance (0.7\%)  and radiative corrections ($<$1\%), yields
 a systematic uncertainty  in $R_D$ of  $\pm$0.028.   At present, there is no reliable
 calculation of   two-photon exchange contributions to the radiative corrections. The two photon
 exchange contribution  to the cross section for a point like proton has been estimated\cite{twophoton} to be
 +0.3\% at $\epsilon=1$ and +2.3\% at  $\epsilon=0$.  If we assume that the contributions for inelastic
 scattering from point like quarks is similar to a point like proton, the two-photon contribution changes $F_{2D}$ by  +0.3\% and $R_D$ by +0.04.  
   
 For targets with atomic number $Z>1$ these  coulomb corrections \cite{Vahe} 
  account for the effect of the  electric field of the additional protons on
  the incident and scattered electrons.  The additional
  protons create an electrostatic potential $V(r)= -\frac{3\alpha(Z-1)}{2R}+\frac{r \alpha(Z-1)}{2R^2}$
  ($R=1.1 A^{1/3}+0.775 A^{-1/3}$  for  atomic weight A).  It has
  been shown\cite{EMA}  that in  the effective
  momentum approximation (EMA) the  
   effective potential   $V_{eff}=0.8 V(r=0)$
  yields results which are in  good agreement
with  the full DWBA calculation\cite{Jin}.   
  This value for $V_{eff}$ also agrees with values extracted from a comparison
  of positron and electron quasi elastic scattering cross sections\cite{Veff}.  
  The  effective incident energy is $E_{eff}=E_0+V_{eff}$,
  and the effective scattered energy is $E'_{eff}= E'+V_{eff}$.  In addition, there is a focusing factor $F_{foc}=\frac{E_0+V_{eff}}{E_0}$. 
  A model of the differential cross section ($\sigma_{model}$) is used to correct the 
  measured cross sections $\sigma_{meas}$ and yield a coulomb corrected(CC) 
  $\sigma^{cc}_{meas}$:
  \begin{equation}
\sigma^{cc}_{meas}(E_0,E') = 
   \frac { \sigma_{meas}(E_0,E')  \sigma_{model}(E_0,E') }{\sigma_{model}(E_{eff},E'_{eff})F^2_{foc}}  
\end{equation} 
where the effective $Q^2$ is $Q^2_{eff}= 4 (E_0+ V_{eff})(E'+V_{eff}) sin^2 (\theta/2).$
	 \begin{figure}[ht]
\includegraphics[width=5.5in,height=4.0in]{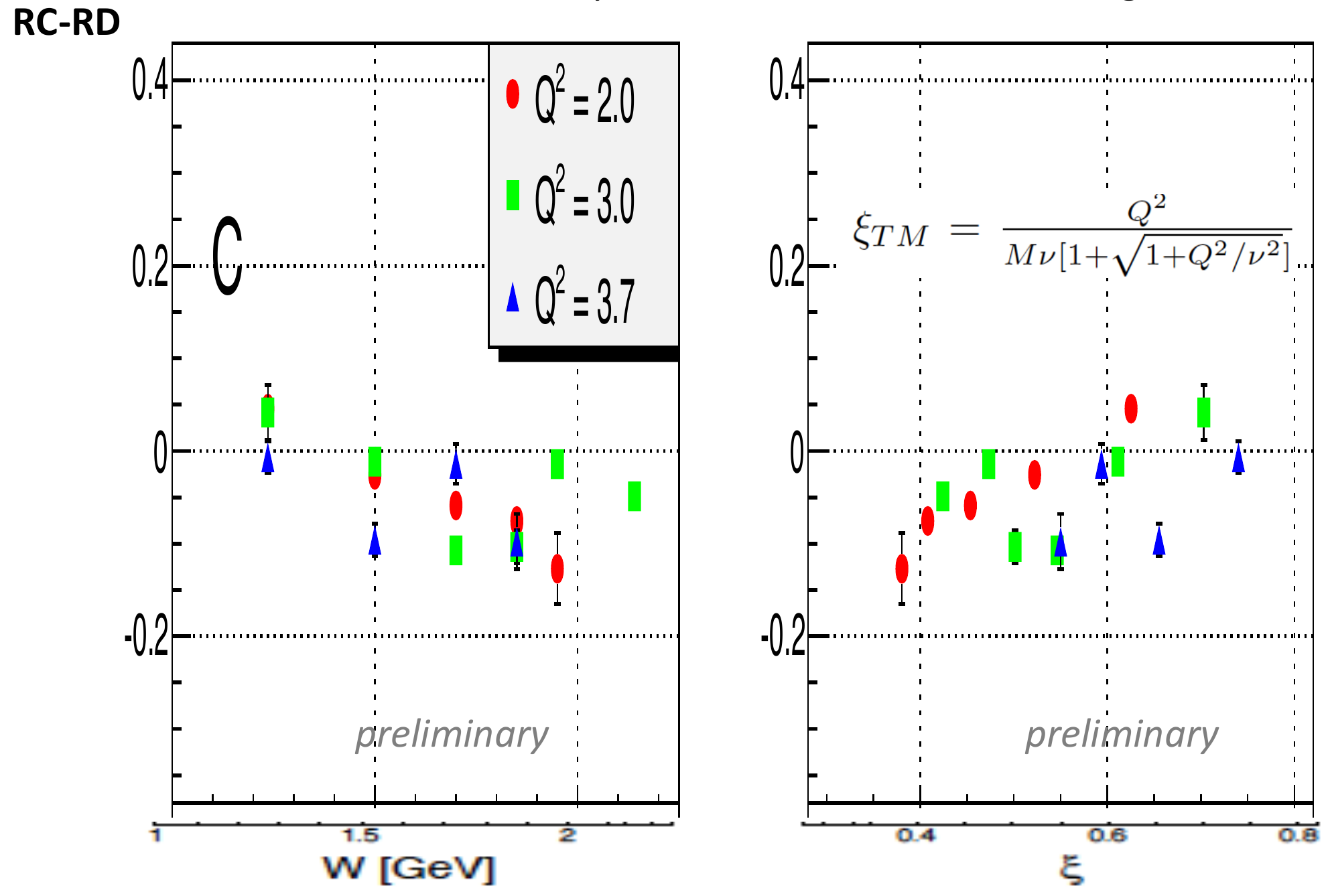}
\caption{ Extracted values of $R_A-R_D$ for
 carbon versus W (left panel) and versus $\xi_{TM}$ (right panel).
 The three values of $Q^2$ are shown as different symbols. 
 Statistical errors only.  
 On average  $R_C< R_D$,  which is contrary to expectation from several theoretical models.
 }
\label{Fig2}
\end{figure}

The difference  $R_D-R_A$ and  $F_{2D}/F_{2A}$ are extracted from linear fits
to the ratio of differential cross section.
\begin{eqnarray}
\frac{\sigma_D}{\sigma_A} = \frac{\sigma_D^T}{\sigma_A^T} 
 \left[1 + \epsilon'(R_D-R_A) \right],\label{req2}
\end{eqnarray} 
where $\epsilon'   = \epsilon /(1+\epsilon R_A)$.   Note that 
$\epsilon R_A$ is small and the resonance structure in $R_A$ is smeared
by Fermi motion.  Therefore,  the extracted $R_D-R_A$ values
are insensitive to the most of the systematic errors listed above
for $R_D$. 
For this study, we bin the data in wide bins of $W$. 

%
%
 
 Fig.~\ref{Fig2} shows the results expressed in terms of  $R_C-R_D$  for
 Carbon  
 versus $W$ (left panel) and versus $\xi_{TM}$ (right panel).
 The three values of $Q^2$ are shown in different symbols.  
 For $W>$ 1.4 GeV  the average value for   $Q^2$= 2, 3, and 4 GeV$^2$   
 is: $\langle R_C-R_D \rangle$= -0.047 $\pm$0.006 (statistical error only).
  
 
 One possible source of systematic error in $R_A-R_D$ is two-photon exchange where the two  photons scatter from quarks in different nucleons in the nucleus.  Some, but not all of the  contribution from 
interactions involving one hard photon absorbed by  one nucleon and multiple soft photons  absorbed
by other nucleons in the nucleus are  included in the Coulomb correction. In addition, the  exchange of
one hard photon with a quark in one nucleon, and another hard photon with a quark in another nucleon is
not included. Theoretical estimates of the contribution of this process are not presently available.
 
Our result that  $R_A<R_D$ is  contrary to several theoretical expectations.  Calculations  \cite{Rfermi}  of the effect of the Fermi motion on nucleons in nuclei predict a small difference in the
opposite direction to what we observe. 
A decrease of the gluon distributions 
in nuclei yields $R_A<R_D$ in the $x<0.1$ shadowing region.  However, the contribution of the gluon distributions to $R$ for  $x>0.3$ is small and these models\cite{gluon} predict  $R_A=R_D$ for the $x$ region of our measurements (i.e. the EMC effect region).
Models\cite{pion} which attribute the EMC effect to the presence of pions in
 nuclei  also predict an effect in the opposite direction to what we observe 
 (these  models also  predict enhancement of antiquarks in the nucleus which has been ruled out by Drell-Yan experiments on nuclear targets\cite{miller}). 
 
 In the naive quark parton model\cite{RF},  $R= 4\langle K_T^2 \rangle /Q^2$, where $K_T$ is the  momentum of quarks in a nucleon perpendicular to the direction of the momentum transfer vector.   Therefore, a possible interpretation of the data is that   $\langle K_T^2 \rangle$ is smaller for  bound nucleons, and that the distributions for both the parallel and perpendicular momentum components of quarks in a nucleon are softened in a nuclear medium  due to partial deconfinement\cite{swell} of quarks caused by short range nucleon-nucleon correlations (SRC) in nuclei\cite{src}.   

In summary,  our preliminary  measurements of    R= $\sigma_L / \sigma_T$ for Carbon ($R_C$) and for deuterium ($R_D$) indicate that  nuclear modifications of the  structure functions of bound nucleons are different for the longitudinal and transverse structure functions, and that  contrary to expectation from several theoretical models,   $R_C< R_D$.



\begin{thebibliography}{99}
\bibitem{swell}  Partial deconfinement of quarks in low momentum nucleons in nuclei has been ruled out
 by  measurements of the form factors of bound nucleons, but not for  high momentum 
 nucleons with high virtuality from SRC.
\end{thebibliography}
\end{document}